\documentclass[12pt]{iopart}
\begin{document}
\title[Stochastic resonance as a filter]{Stochastic resonance as a filter for signal 
detection from multi signal inputs}

\author{K P Harikrishnan\dag and G.Ambika\ddag}

\address{\dag\ Department of Physics, The Cochin College, Cochin-682002, India}
\ead{$kp_{\_}hk2002@yahoo.co.in$}
\address{\ddag\ Department of Physics, Maharajas College, Cochin-682011, India}
\ead{$gambika4nk@yahoo.com$}

\begin{abstract}
We undertake a detailed numerical study of the phenomenon of stochastic 
resonance with multisignal inputs. A bistable cubic map is used as the model 
and we show that it combines 
the features of a bistable system and a threshold system. A study of 
stochastic 
resonance in these two setups reveal some fundamental differences between 
the two mechanisms with respect to amplification of a composite input signal. 
As a practically relevant result, we show that the phenomenon of stochastic 
resonance can be 
used as a \emph{filter} for the detection/transmission of the component 
frequencies in a composite signal.
\end{abstract}


\maketitle

\section{Introduction}

Stochastic Resonance(SR) refers to the situation wherein the response of a 
nonlinear system to a weak input signal can be significantly increased with 
appropriate tuning of the noise intensity[1,2]. When a subthreshold signal 
$I(t)$ is input to a nonlinear system $g$ together with a noise 
$\zeta (t)$, if the filtered output $O(t) \equiv g(I(t)+ \zeta (t))$ 
shows enhanced response that contains the information content of $I(t)$, 
then SR is said to be realised in the system. The mechanism first used by 
Benzi, Nicolis etc [3,4] to explain natural phenomena is now being used 
for a large variety of interesting applications like modelling biological 
and ecological 
systems [5], lossless communication purposes etc [6]. Apart from these, it 
has opened up a vista of many related resonance phenomena [7] which are 
equally challenging from the point of view of intense research. In this work, 
we try to capture some of these, using a simple model system, namely, a two 
parameter bimodal cubic map.

In the early stages of the development of SR, most of the studies were done 
using a dynamical set up with bistability, modelled by a double well 
potential. Here SR is realised due to the shuttling between the two stable 
states at the frequency of the subthreshold signal with the help of noise. 
Thus if the potential is 
\begin{equation}
V(X) = -a X^2/2 + b X^4/4
\label{eqn1}
\end{equation}
in the presence of a signal and noise, the dynamics can be modelled by an 
overdamped oscillator
\begin{equation}
\dot{X} = a X - b X^3 + ZSin \omega t + E \zeta(t)
\label{eqn2}
\end{equation}
If $C_{th}$ is the threshold at which deterministic switching (with noise 
amplitude $E=0$) is 
possible, then well to well switching due to SR occurs when
\begin{equation}
-C_{th} > (ZSin \omega t+E \zeta(t)) > C_{th}
\label{eqn3}
\end{equation}
that is, twice in one period of the signal $T = 2 \pi/\omega$.

The characterisation of SR in this case is most commonly done by computing 
the signal to noise ratio (SNR) from the power spectrum of the output as
\begin{equation}
SNR = 10log_{10} (S/N)  dB
\label{eqn4}
\end{equation}
where $N$ is the average background noise around the signal $S$.
If SR occurs in the system, then the SNR goes through a maximum giving a bell 
shaped curve as $E$ is tuned.

SR has also been observed in systems with a single stable state with an 
escape scenerio. These \emph{threshold} systems, in the simplest case, can 
be modelled by a piecewise linear system or step function

$$g(u) = -1 \hspace {12pt} u < C_{th}$$\\
$$\hspace {22pt} = +1 \hspace {12pt} u > C_{th}$$\\
The escape with the help of noise is followed by resetting to the monostable 
state. In this case, a quantitative characterisation is possible directly 
from the output, but only in terms of probabilities. If $t_n$ are the escape 
times, the inter spike interval (ISI) is defined as $T_n = t_{n+1}-t_n$ and 
$m(T_n)$ is the number of times the same $T_n$ occurs. For SR to be realised 
in the system, the probability $p_n = m(T_n)/N$ ($N$ is the total number of 
escapes) has to be maximum corresponding to the signal period $T$ at an 
optimum noise amplitude.

There are situations where SR is to be optimised by adapting to or 
designing the dynamics of the system. This is especially relevant in natural 
systems or electronic circuits where the noise is mostly from the 
environmental background and therefore not viable to fine tuning. Similarly, 
depending on the context or application, the nature of the signal can also 
be different, such as, periodic, aperiodic, digital, composite etc. The 
classical SR deals with the detection of a single subthreshold signal 
immersed in noise. However, in many practical situations, a composite signal 
consisting of two or more harmonic components in the presence of 
background noise is encountered, as for example, in biological systems for the  
study of planktons and human visual cortex [8], in laser physics [9] and in 
accoustics [10]. Moreover, two frequency signals are commonly used in 
multichannel optical communication systems based on wave length division 
multiplexing (WDM) [11]. But very few studies of SR have been carried out 
involving such \emph{bichromatic} signals to date [12-15], and each of them 
pertaining to some specific dynamical set ups. This motivates us to 
undertake a detailed numerical analysis of SR with such signals using a 
simple model, a two parameter cubic map. It is a discretised version of the 
overdamped bistable oscillator, but with the added feature of an inherent 
escape mode also. Hence it can function in both set ups, bistable and 
threshold, as a stochastic resonator.

Our analysis brings out some generic features (and also a few interesting 
differences) of both these mechanisms of SR with respect to multisignal 
inputs. An especially important result that has emerged from our studies 
is that SR can, in principle, be used as a \emph{filter} to detect the 
fundamental frequencies in a composite signal by tuning the noise 
amplitude. Our paper is organised as      
follows: In $\S 2$, we introduce the cubic map and discuss some theoretical 
aspects of SR with multiple signals. In $\S 3$, we study SR in the system 
numerically with a composite signal treating it as a bistable system. 
$\S 4$ considers the cubic map as a threshold system having properties 
different from that of a bistable system. Results and discussions are 
given in $\S 5$.

\input psbox.tex

\begin{figure}
\begin{center}
{\mbox{\psboxto(14cm;16cm){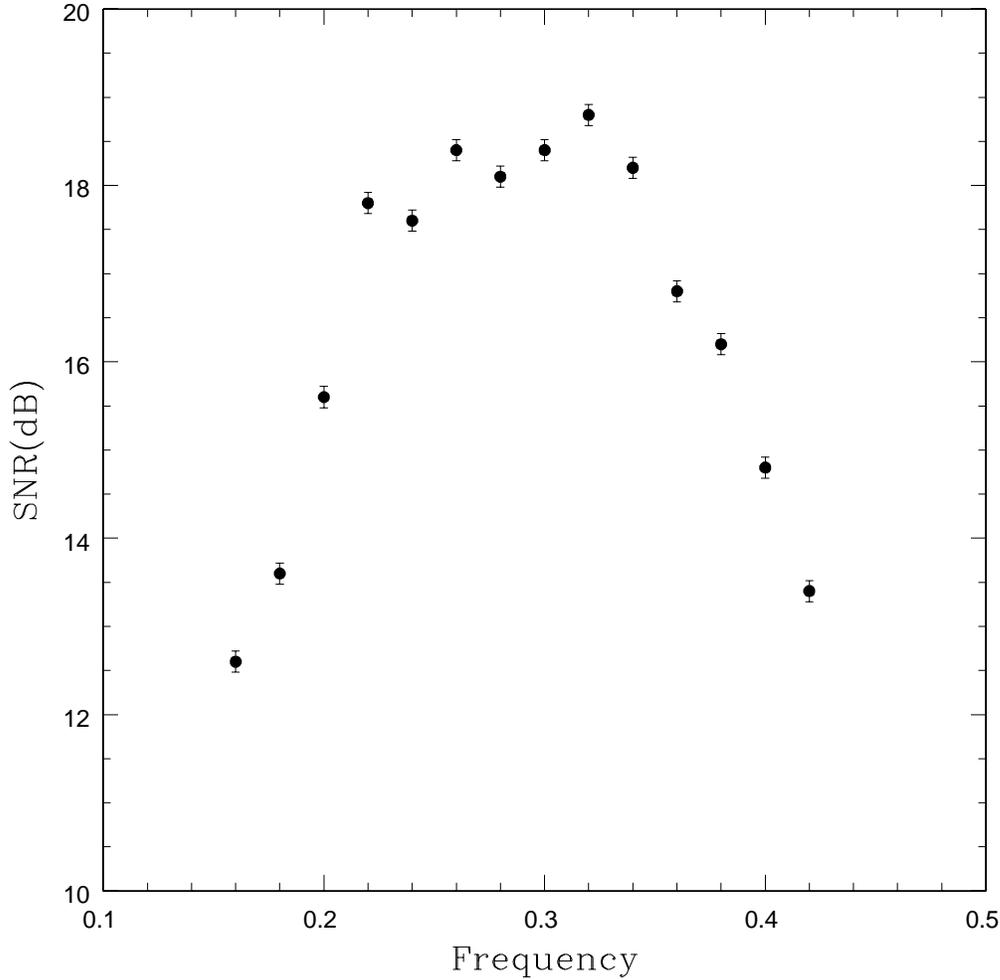}}}
\end{center}
\caption{Variation of SNR with frequency for the bistable cubic 
map in the 
chaotic regime $(a=2.4)$ with the noise amplitude $E=0$. Note that a 
subthreshold signal $(Z=0.16)$ can be detected for an intermediate range of 
frequencies, without any external noise.}
\label{Fig.1label}
\end{figure}

\section{The model}
The two parameter cubic map is given by
\begin{equation}
X_{n+1}= f(X_{n})= b + a X_{n} - X_{n}^3
\label{eqn5}
\end{equation}
The system has been studied in great detail both analytically and numerically 
and has been shown to possess a 
rich variety of dynamical properties including bistability [16]. In particular,
if $a_1$ is the value of the parameter at which $f^{\prime}(X_i,a_1,b) =1$,
then for 
$a>a_1$, there is a window in $b$, where bistability is observed. The bistable 
attractors are clearly separated with $X>0$ being the basin of one and 
$X<0$ that of the other. For example, for $a=1.4$, $b=[-0.1,0.1]$, two 
attractors 
of period 1 coexist. As the value of $a$ is increased, the periodicity of the 
bistable attractors keeps on doubling while the width of the window around 
$b$ decrease correspondingly. Finally, for $a=2.4$, two chaotic attractors 
co-exist in a very narrow window around $b$.

The system when driven by a gaussian noise and a periodic signal becomes 
\begin{equation}
X_{n+1} = b + a X_n - X_{n}^3 + E \zeta(t) + Z F(t)
\label{eqn6}
\end{equation}
where we choose $\zeta(t)$ to be a gaussian noise with zero mean and $F(t)$ is the 
periodic signal. The amplitude of the noise and the signal can be varied by 
tuning E and Z respectively. It can be shown that in the regime of chaotic 
bistable attractors $(a=2.4,b=0.01)$, a subthreshold input signal can be 
detected using the inherent chaos in 
the system without any external noise(E=0). Taking the signal 
$F(t) = Z Sin(2 \pi \nu t)$, with $Z = 0.16$, the system shows SR type 
behavior 
for an optimum range of frequencies as shown in Fig.1, where the output SNR 
is plotted 
as a function of the frequency $\nu$. It implies that a subthreshold signal  
can be detected by passing through a bistable system making use of the 
inherent chaos in it without the help of any external noise. This 
phenomenon is known as \emph {deterministic resonance}. In the regime of 
periodic bistable attractors $(a=1.4,b=0.01)$ with a single subthreshold 
signal, the system shows conventional SR as well as chaotic resonance (CR), 
and using this model, we 
have recently reported some new results including enhancement of SNR via 
coupling [17].

\begin{figure}
\begin{center}
{\mbox{\psboxto(14cm;16cm){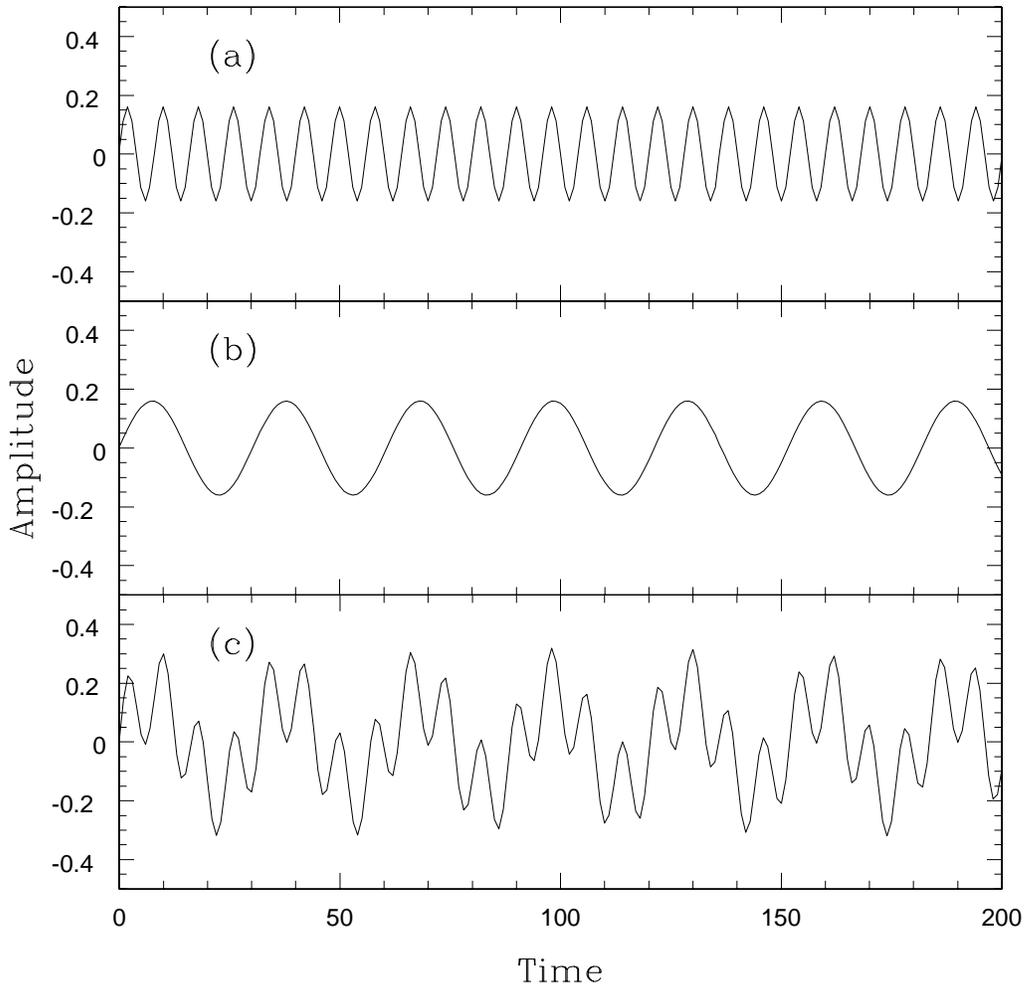}}}
\end{center}
\caption{Time variation of a signal with frequency $(a) \nu_1=0.125$ and 
$(b) \nu_2=0.033$, with amplitude $Z=0.16$ in both cases. When these signals 
are superposed, the resulting signal is shown in fig.(c), which is in 
accordance with the linear superposition principle.}
\label{Fig.2label}
\end{figure}
 
For the remaining part of the paper we consider the signal $F(t)$ to be a 
combination of at least 2 frequencies, $\nu_1$ and $\nu_2$. A typical input 
signal is shown in Fig.2(c) which is a superposition of two fundamental 
frequencies $\nu_1=0.125$ and $\nu_2=0.033$ shown in Fig.2(a) and 2(b) 
respectively. Before we go into 
the numerical studies, we consider some theoretical results for multisignal 
inputs. An analytical description of SR usually considers the model of an 
overdamped bistable oscillator in a double well potential, driven by white 
noise $\zeta(t)$ and a periodic signal $F(t) = Z Sin(2 \pi \nu t)$. An 
expression 
for the SNR can be derived using some approximate theories, the most popular 
being the Linear Response Theory(LRT) [18-21].

According to this theory, the response of a nonlinear stochastic system 
$X(t)$ to a weak external force $F(t)$ in the asymptotic limit of large 
times is determined by the integral relation [18]
\begin{equation}
X(t) = <X_0> + \int_{-\infty}^{\infty} R(t-\tau) F(\tau) d \tau
\label{eqn7}
\end{equation}
where $<X_0>$ is the mean value of the state variable for $F(t)=0$
Without lack of generality, one can set $<X_0>=0$. The function $R(t)$ in 
(7) is called the response function. For a harmonic signal, the system response 
can be expressed through the function $R(\omega)$ which is the Fourier 
transform of the response function:
\begin{equation}
X(t) = Z |R(\omega)| Sin(2 \pi \nu t + \psi)
\label{eqn8}
\end{equation}
where $\psi$ is a phase shift.

The LRT can be naturally extended to the case of multifrequency signals. Let 
the signal $F(t)$ be a composite signal of the form:
\begin{equation}
F(t) = Z \sum_{k=1}^{n} Sin(2 \pi \nu_{k}t)
\label{eqn9}
\end{equation}
where $\nu_{k}$ are the frequencies of the discrete spectral components with 
the same amplitude $Z$. Then according to LRT, the system response can be 
shown to be [21]
\begin{equation}
X(t) = Z \sum_{k=1}^{n} |R(\omega_{k})| Sin(2 \pi \nu_{k}t + \psi_{k})
\label{eqn10}
\end{equation}
which contains the same spectral components at the input, but with different 
amplitudes and phases. We now investigate this numerically in more detail 
using system (5), both as a bistable system and a threshold system. 

\section{The cubic map as a bistable system}

\begin{figure}
\begin{center}
{\mbox{\psboxto(14cm;16cm){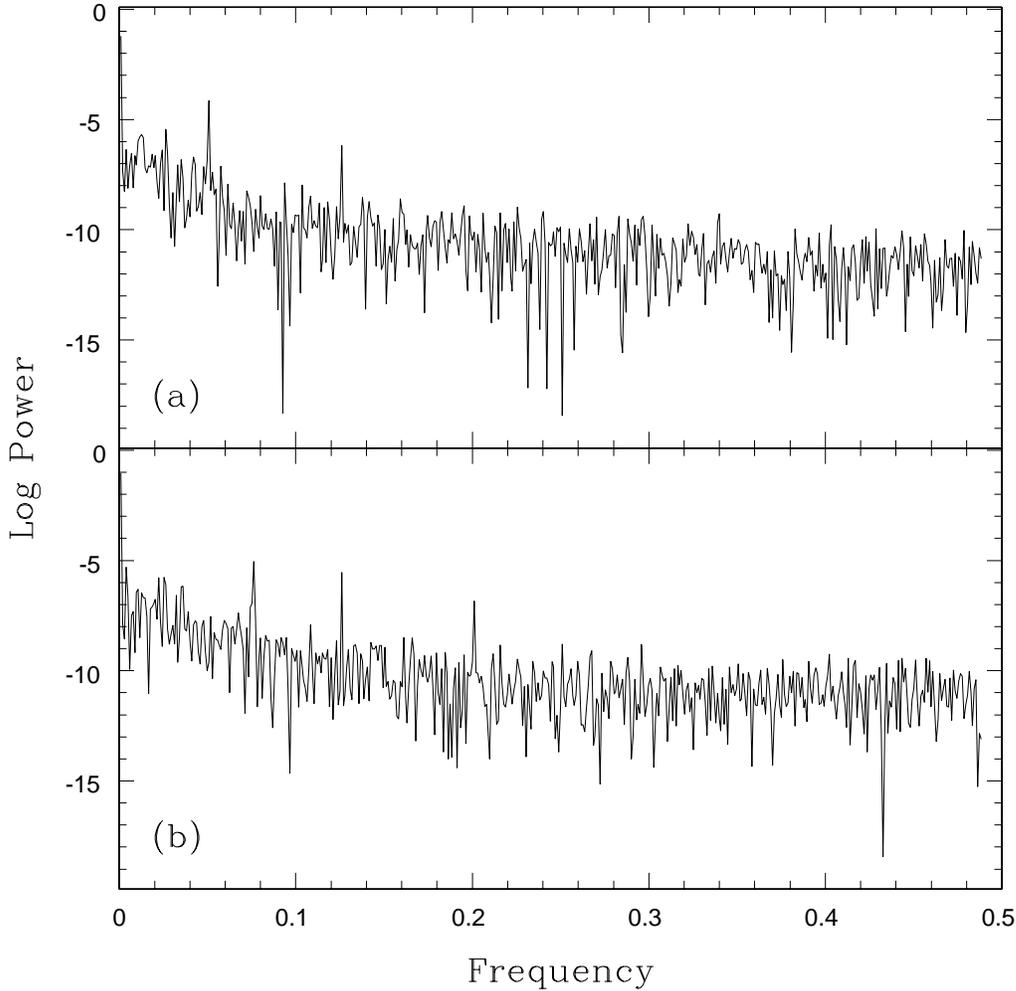}}}
\end{center}
\caption{Power spectrum for the bistable system (6) with a composite signal 
consisting of $(a)$ 2 frequencies $(\nu_1,\nu_2) = (0.125,0.05)$ and $(b)$ 3 
frequencies $(\nu_1,\nu_2,\nu_3) = (0.125,0.075,0.2)$, with $Z=0.16$ and 
$E=0.4$ in both cases.}
\label{Fig.3label}
\end{figure}

Taking $F(t)$ as a composite signal as in equation (9) with $Z=0.16$ for 
subthreshold 
signals of equal amplitude, we drive the system using different combination of 
frequencies $\nu_{k}$, with $a=1.4$ and $b=0.01$ in the regime of bistable 
periodic attractors. For convenience, $\nu_{1}$ is fixed as 0.125 and the 
other 
frequencies $\nu_{2}, \nu_{3}$ etc are varied from 0.02 to 0.4 in steps of 
0.005. For each selected combination, the output power spectrum is calculated 
using the FFT algorithm, for different values of the noise amplitude $E$. Two 
typical power 
spectrums for n=2 and n=3 with $(\nu_{1},\nu_{2})=(0.125,0.05)$ and 
$(\nu_{1},\nu_{2},\nu_{3})=(0.125,0.075,0.2)$ are shown in Fig.3(a) and (b) 
respectively. To compute the power spectrum, only the inter-well transitions 
are taken into account and all the intra-well fluctuations are suppressed 
with a two state filtering. It is clear that only the fundamental frequencies 
present in the input are enhanced.

\begin{figure}
\begin{center}
{\mbox{\psboxto(14cm;16cm){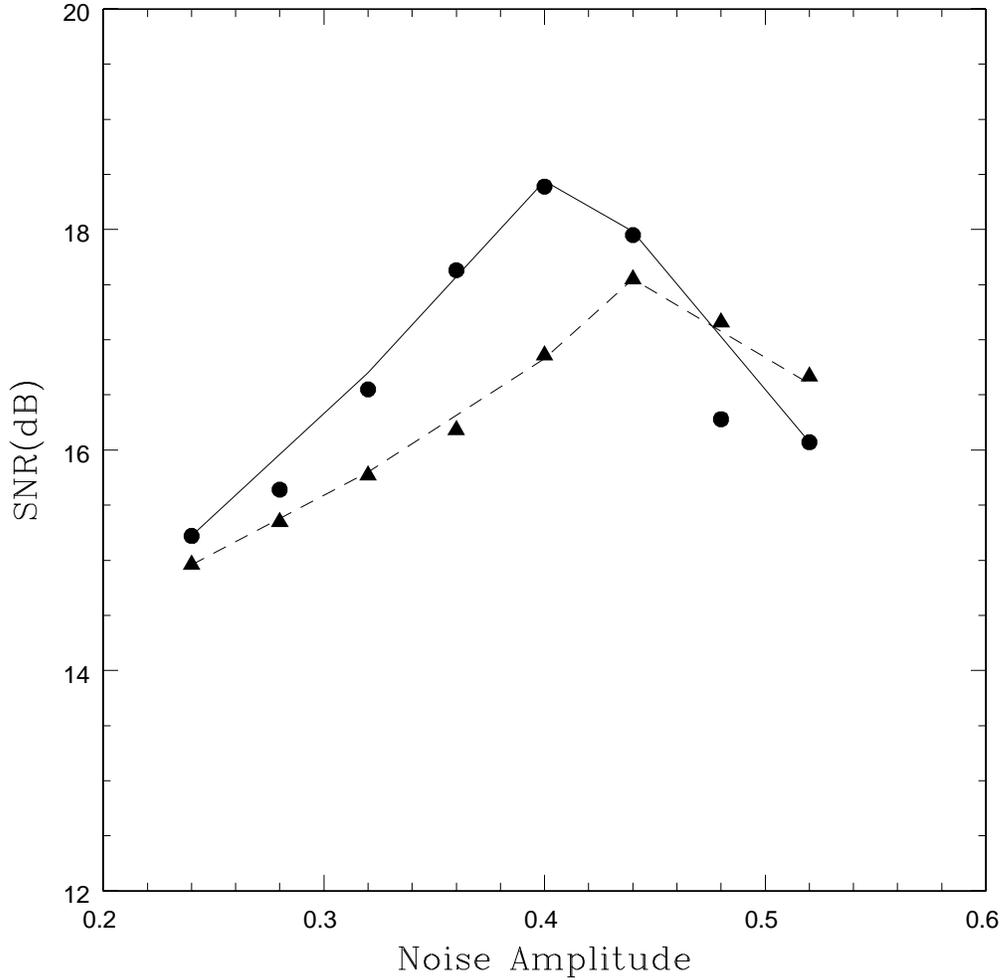}}}
\end{center}
\caption{Variation of SNR with noise amplitude for the 2 frequencies shown in 
$Fig.3(a)$, with the filled circle representing the higher frequency $\nu_1$.}
\label{Fig.4label}
\end{figure}

\begin{figure}
\begin{center}
{\mbox{\psboxto(14cm;16cm){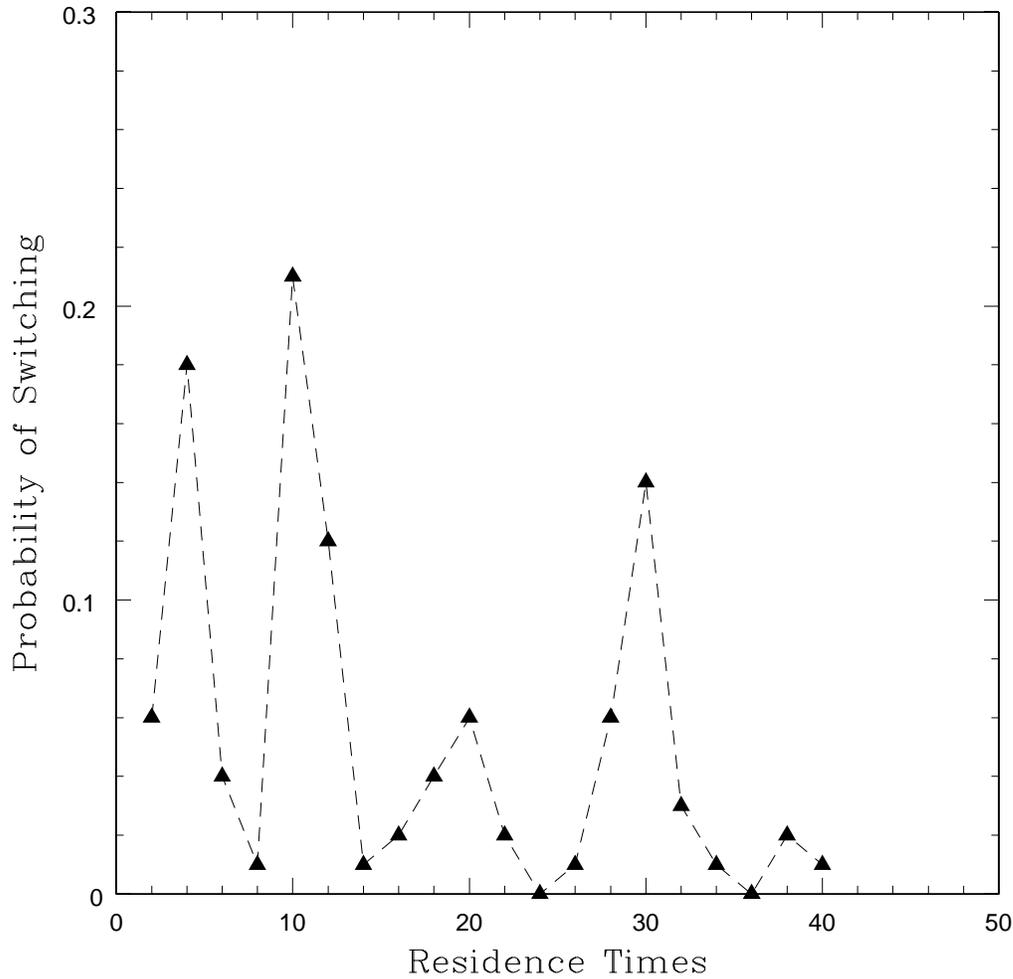}}}
\end{center}
\caption{The RTDF for the system (6),with a composite signal consisting of 2 
frequencies $\nu_1=0.125$ and $\nu_2=0.05$. Note that the peaks are 
synchronised with half the periods corresponding to these frequencies.}
\label{Fig.5label}
\end{figure}

We now concentrate on n=2 (a combination of 2 frequencies) and compute the 
two important quantifiers of SR, namely, the SNR and the Residence Time 
Distribution Function(RTDF). For the frequencies in Fig.3(a), the SNR is 
computed from the power spectrum using equation (4)
for a range of values of $E$ and the results are shown in Fig.4. 
The RTDF measures the probability distribution of the average times the 
system resides in one basin, as a function of different periods. If T is the 
period of the applied signal, the distribution will have peaks corresponding 
to times $(2n+1)T/2$, n=0,1,2...For the system (6) with $(\nu_{1},\nu_{2}) = 
(0.125,0.05)$, the results are shown in Fig.5. Note that there are only peaks 
corresponding to the half integer periods of the two applied frequencies.

The above computations are repeated taking various combinations of 
frequencies $(\nu_1,\nu_2)$, both commensurate and non-commensurate.  
For a fixed combination of $(\nu_{1},\nu_{2})$, the calculations  
are also done by changing the signal amplitude $Z$ of one and both signals. 
Always the results remain qualitatively the same and only the fundamental 
frequencies 
present in the input are amplified at the output. If $Z$ becomes very small
$(< 0.1)$ compared to the noise level, then the phenomenon of SR disappears 
altogether and the signal remains undetected in the background noise.

\begin{figure}
\begin{center}
{\mbox{\psboxto(14cm;16cm){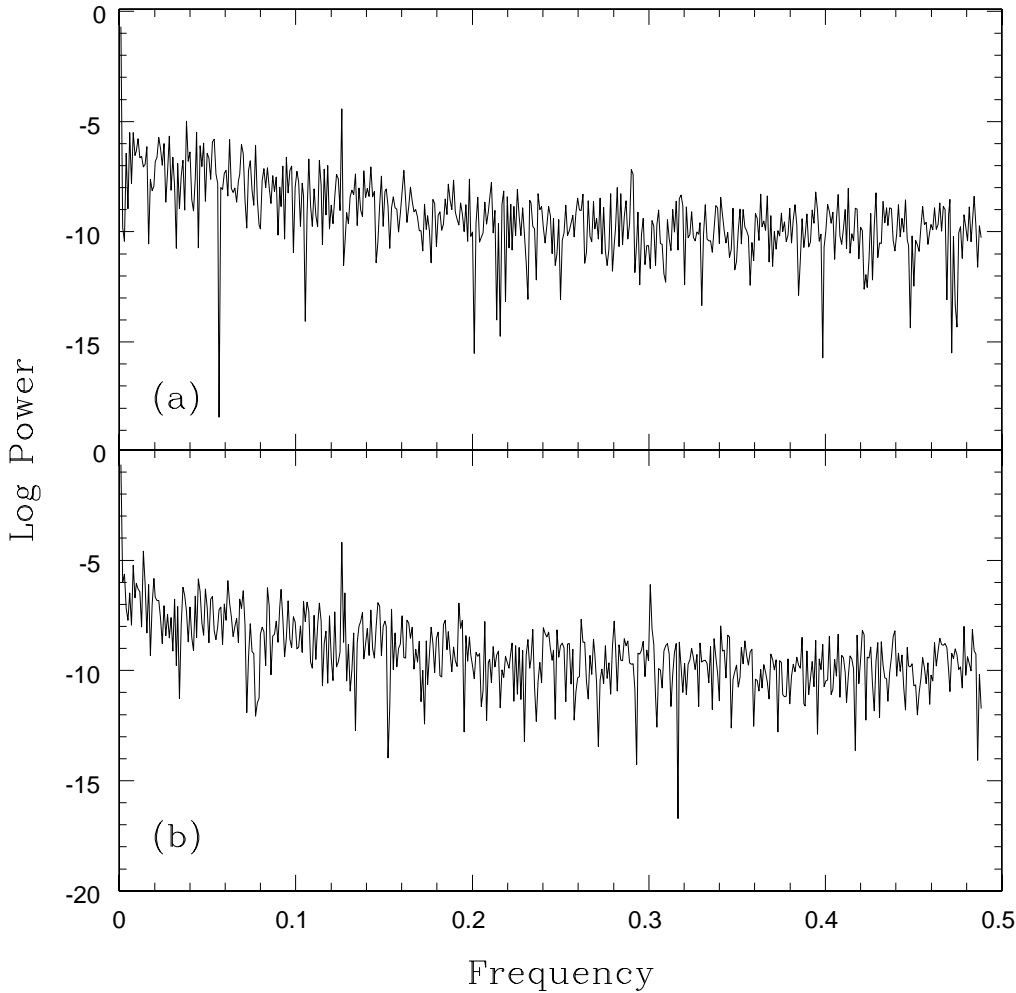}}}
\end{center}
\caption{The power spectrum for the bistable system (11), with the signal 
$F(t)$ consisting of $(a)$ one frequency $\nu_1=0.125$ and $(b)$ 2 frequencies 
$\nu_1=0.125$ and $\nu_2=0.3$. The parameter values used are $Z=0.16, E=1.0, 
a=1.4$ and $b=0.01$.}
\label{Fig.6label}
\end{figure}

In all the above computations, we used \emph{additive} noise, where the noise 
has been added to the system externally. But in many natural systems, noise 
enters through an interaction of the system with the surroundings, that is,
through a parameter modulation, rather than a simple addition. Such a 
\emph{multiplicative} noise occurs in a variety of physical phenomena [22] 
and can, in principle, show qualitatively different behavior in the presence 
of a periodic field [23]. To study its effect on the bistable 
system, equation (6) is modified as
\begin{equation}
X_{n+1} = b + a(1+E \zeta (t))X_{n} - X_{n}^3 + Z F(t)
\label{eqn11}
\end{equation}
The noise is added through the parameter $a$ which determines the nature of 
the bistable attractors. With $a=1.4$ and $b=0.01$, the system is now driven 
by a signal of single frequency $\nu_{1}=0.125$ and a multisignal with 2 
frequencies $(\nu_{1},\nu_{2})$, with $Z=0.16$. The power spectrum for single 
frequency and multiple frequencies are shown in Fig.6(a) and (b) 
respectively. The corresponding SNR variation with noise $E$ are shown in 
Fig.7(a) and (b). Note that the results are qualitatively identical to 
that of additive noise, but the optimum SNR and the corresponding noise 
amplitude are comparitively much higher in this case. Thus 
our numerical results indicate that a bistable system responds only to the 
fundamental frequencies in a composite signal and not to any mixed modes 
such as $(\nu_{1}-\nu_{2})$, in agreement with the LRT. Next we consider the 
cubic map as a monostable threshold system and show that it behaves 
differently.

\begin{figure}
\begin{center}
{\mbox{\psboxto(14cm;16cm){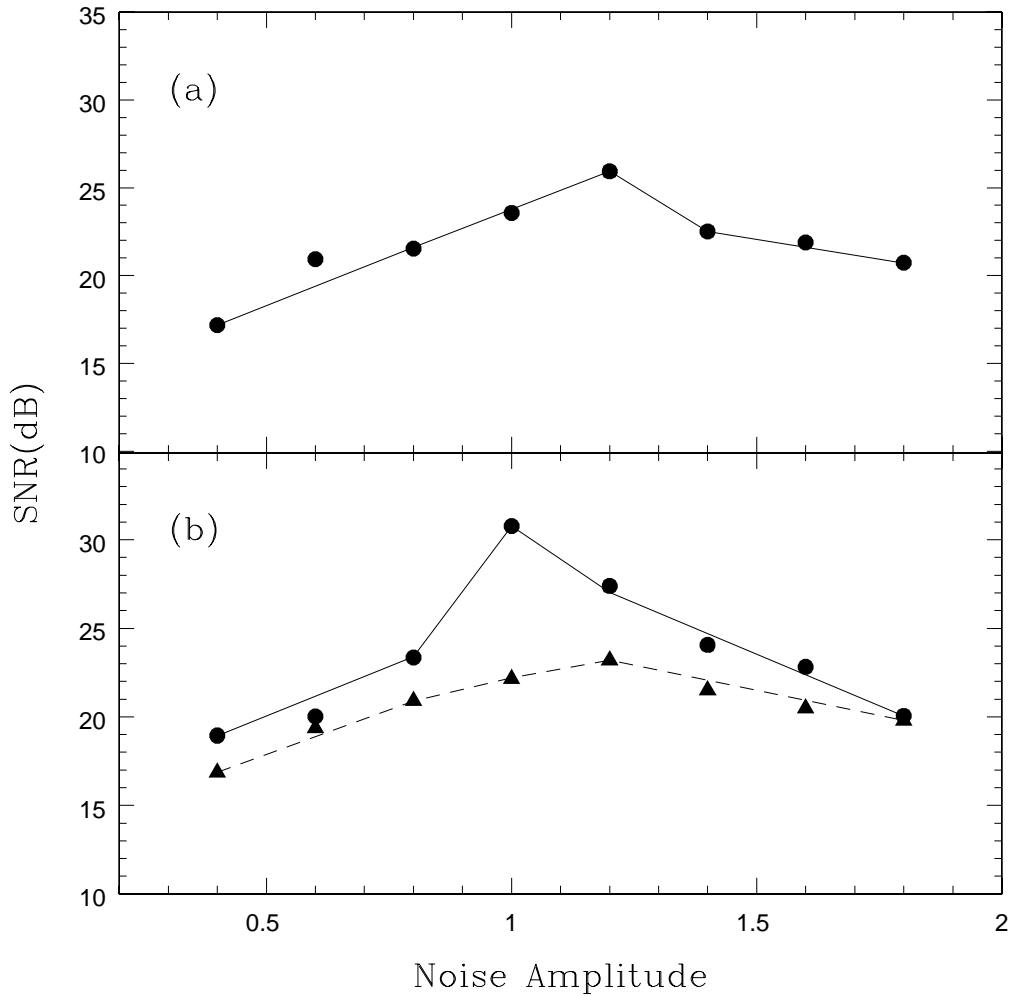}}}
\end{center}
\caption{Variation of SNR with noise for the system(11), for $(a)$ single 
frequency $\nu_1=0.125$ and $(b)$ 2 frequencies $\nu_1=0.125$ (filled 
circles) and $\nu_2=0.3$. Note that the optimum SNR of $\nu_1$ increases by 
about 5 dB when a second signal $\nu_2$ is added.}
\label{Fig.7label}
\end{figure}

\section{The cubic map as a threshold system}
As said earlier, the domains of the bistable attractors in the cubic map are 
clearly separated with the boundary $X=0$. Hence the cubic map can also be 
considered as a nondynamical threshold system with a single stable state 
having a potential barrier. Here the system generates an output \emph{spike} 
only when the combined effort of the signal and the noise pushes it across 
the potential barrier $(at X=0)$ in one direction:
\begin{equation}
[E \zeta (t) + Z F(t)] > C_{th}
\label{eqn12}
\end{equation} 
It is then externally reinjected back into the basin. The output thus 
consists of a series of spikes similar to a random telegraph process. The 
study of SR in such systems assumes importance in the context of biological 
applications and in particular the integrate and fire models of neurons 
where SR has been firmly established[2,24].

The computations are done using equation (6), initially with a single frequency 
signal, $F (t) = Sin(2 \pi \nu_{1} t)$. We start from an initial condition in 
the negative basin and when the output crosses the threshold, $X=0$, it is 
reinjected back into the basin by resetting the initial conditions. This is 
repeated for a sufficiently large number of escapes and the ISIs are 
calculated. The ISIs are then normalised in terms of the 
periods $T_{n}$ and the probability of escape corresponding to each $T_{n}$ 
is calculated as the ratio of the number of times $T_{n}$ occurs to the 
total number of escapes. The whole procedure is repeated tuning the noise 
amplitude $E$. It is found that the ISI is synchronised with the period of the 
forcing signal for an \emph{optimum} noise amplitude (Fig.8), indicating SR 
for the frequency $\nu_{1}$.

\begin{figure}
\begin{center}
{\mbox{\psboxto(14cm;16cm){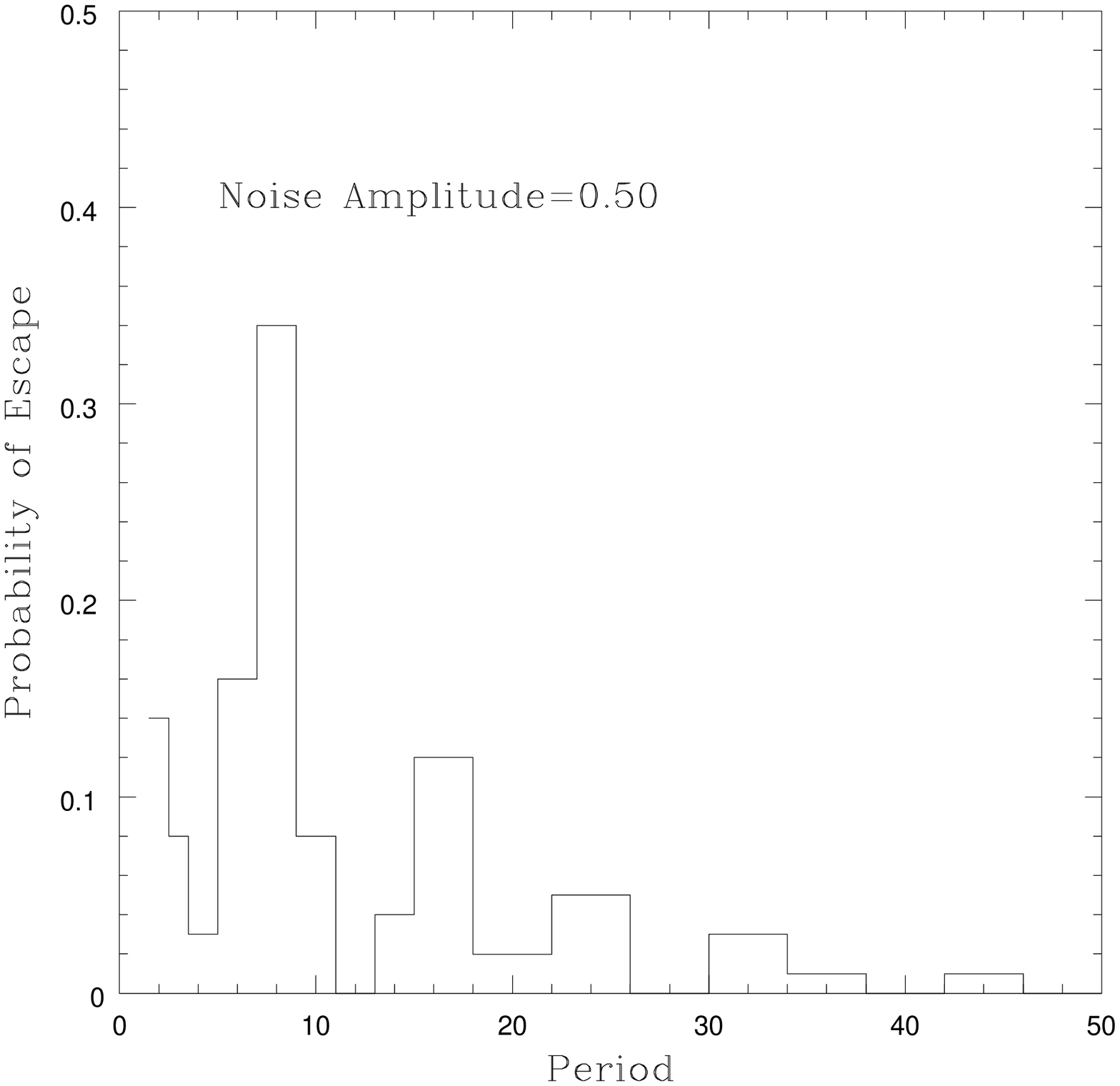}}}
\end{center}
\caption{The probability of escape for the system (6) corresponding to 
different periods, when it is used as a threshold system  
with the signal $F(t)$ having a single frequency $\nu_1=0.125$.} 
\label{Fig.8label}
\end{figure}

The calculations are now repeated by adding a second signal of frequency 
$\nu_{2}$ and amplitude same as that of $\nu_{1}$. Again different values of 
$\nu_{2}$ in the range 0.02 to 0.4 are used for the calculation. It is then 
found that apart from the input frequencies $\nu_{1}$ and $\nu_{2}$, a third 
frequency, which is a mixed mode is also enhanced at the output, at a lesser 
value of the noise amplitude. To make it clear, the amplitude $Z$ of both 
$\nu_{1}$ and $\nu_{2}$ are reduced from 0.16 to 0.08, so that they become 
too weak to get amplified. The results of computations are shown in Fig.9 and 
Fig.10, for a combination of input signals $(\nu_1,\nu_2)=(0.125,0.033)$.
Fig.9 represents the probability of escape corresponding to different periods,
for the optimum value of noise. It is clear that only 
a very narrow band of frequencies $d \nu$ around a third frequency $\approx 0.045$ 
(corresponding to the period $T \approx 22s$), are 
amplified at the output. Note that this frequency is absent in the input and 
corresponds to $(\nu_1 - \nu_2)/2$. This is in sharp contrast to the earlier 
case of a bistable system. The variation of escape probability corresponding to 
this frequency as a function of noise amplitude is shown in Fig.10.

\begin{figure}
\begin{center}
{\mbox{\psboxto(14cm;16cm){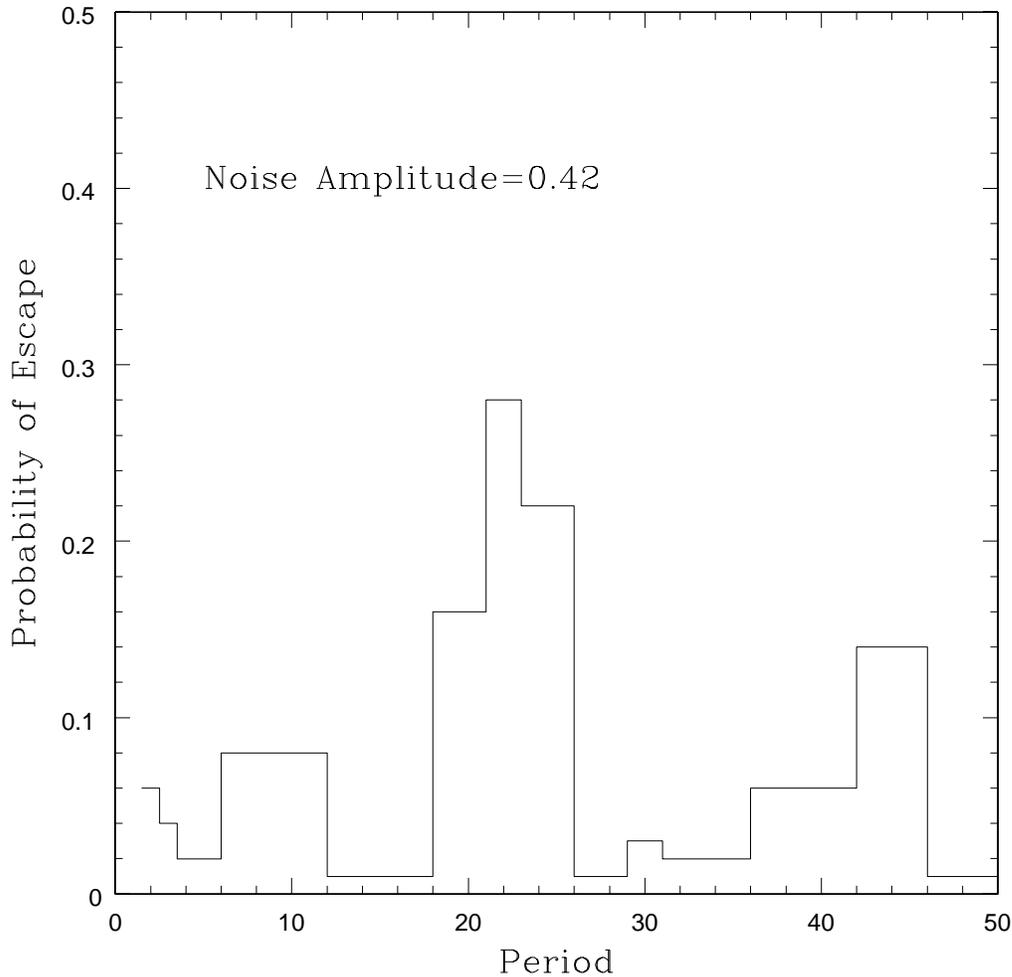}}}
\end{center}
\caption{Same as $Fig.8$, but with 
the signal $F(t)$ comprising of 2 frequencies $\nu_1=0.125$ 
and $\nu_2=0.033$ with the individual amplitudes below that required for SR. 
Note that only a small band of frequencies around 
$(\nu_1 - \nu_2)/2$ are amplified.} 
\label{Fig.9label}
\end{figure}

\begin{figure}
\begin{center}
{\mbox{\psboxto(14cm;16cm){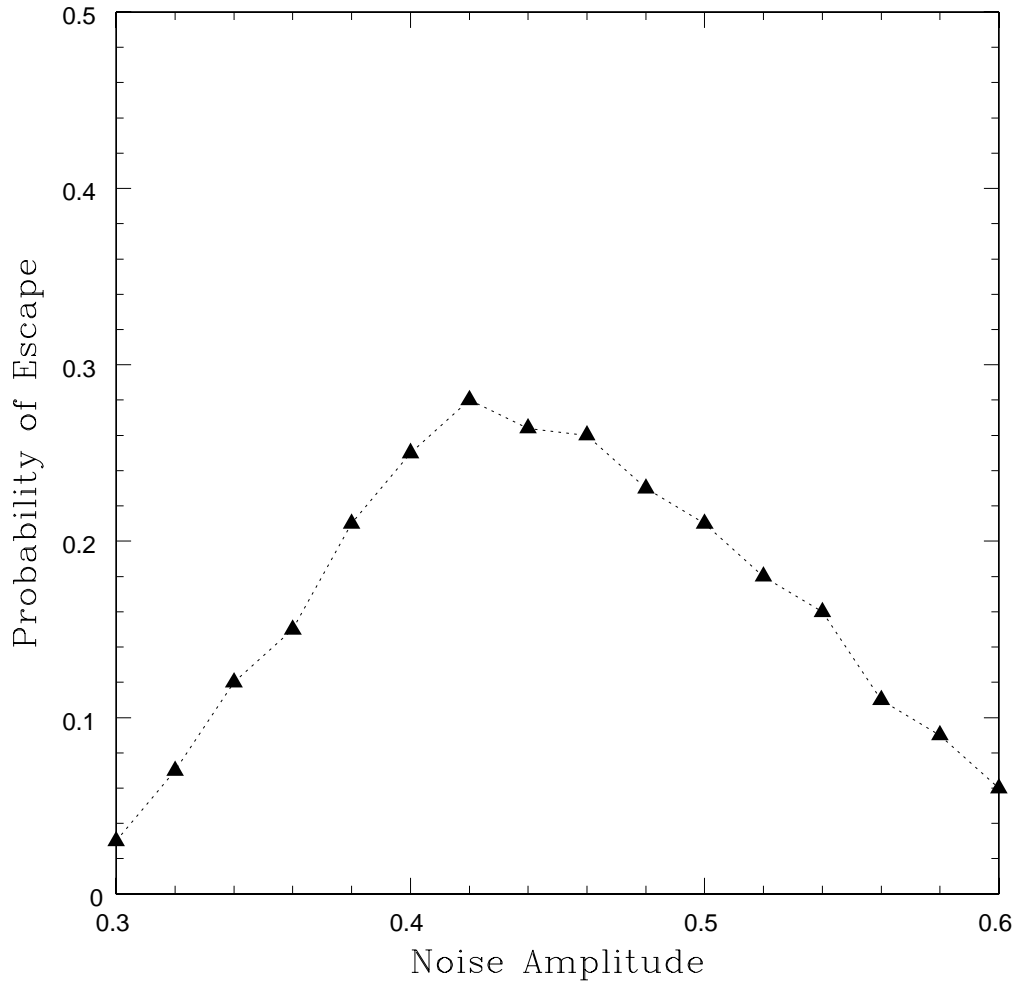}}}
\end{center}
\caption{Variation of the probability of escape corresponding to the frequency 
$(\nu_1 - \nu_2)/2$ for the system (6) as a function of noise amplitude, when it is 
used as a threshold system with an input signal consisting of 2 frequencies 
$\nu_1$ and $\nu_2$. }
\label{Fig.10label}
\end{figure}

\begin{figure}
\begin{center}
{\mbox{\psboxto(14cm;16cm){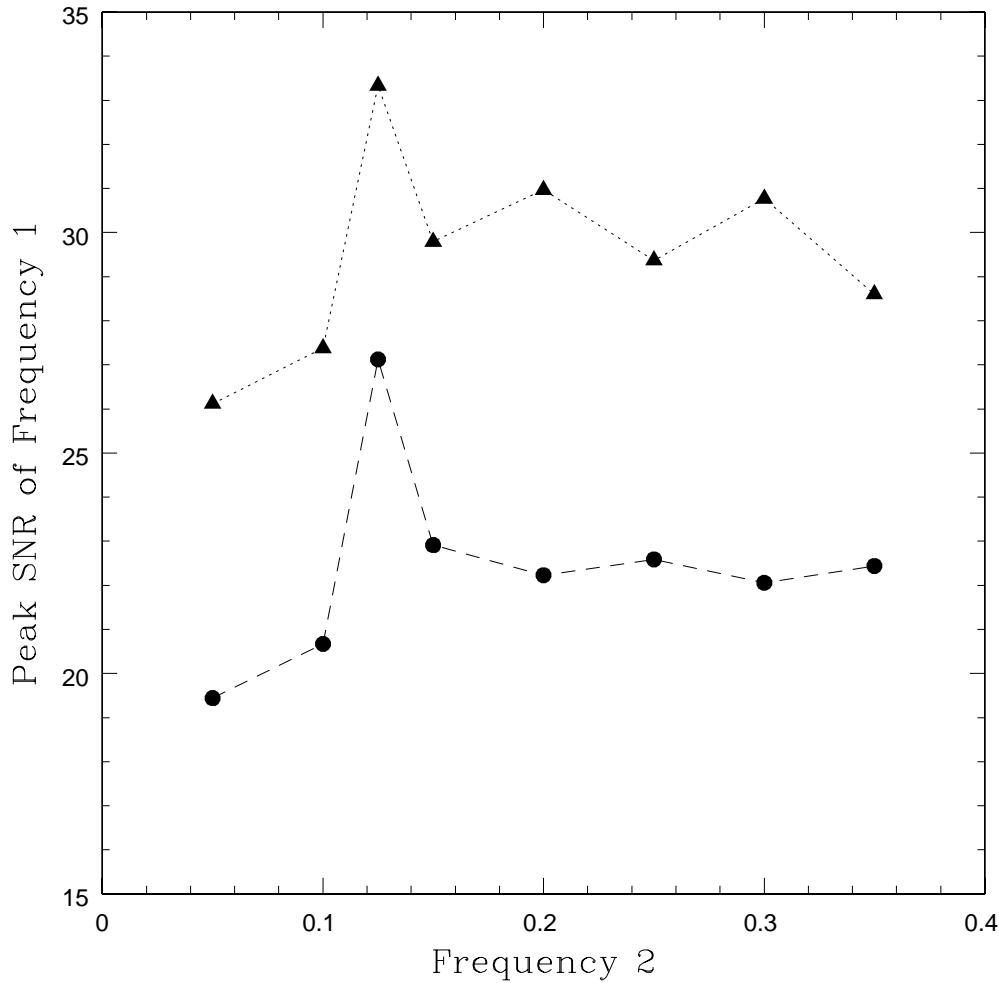}}}
\end{center}
\caption{Variation of the optimum value of SNR for the frequency $\nu_1$ as 
the second frequency $\nu_2$ is varied, for additive noise(filled circles) 
and multiplicative noise(filled triangles). Note that for $\nu_1 = \nu_2$, the 
SNR is considerably enhanced in both cases.}
\label{Fig.11label}
\end{figure}

\begin{figure}
\begin{center}
{\mbox{\psboxto(14cm;16cm){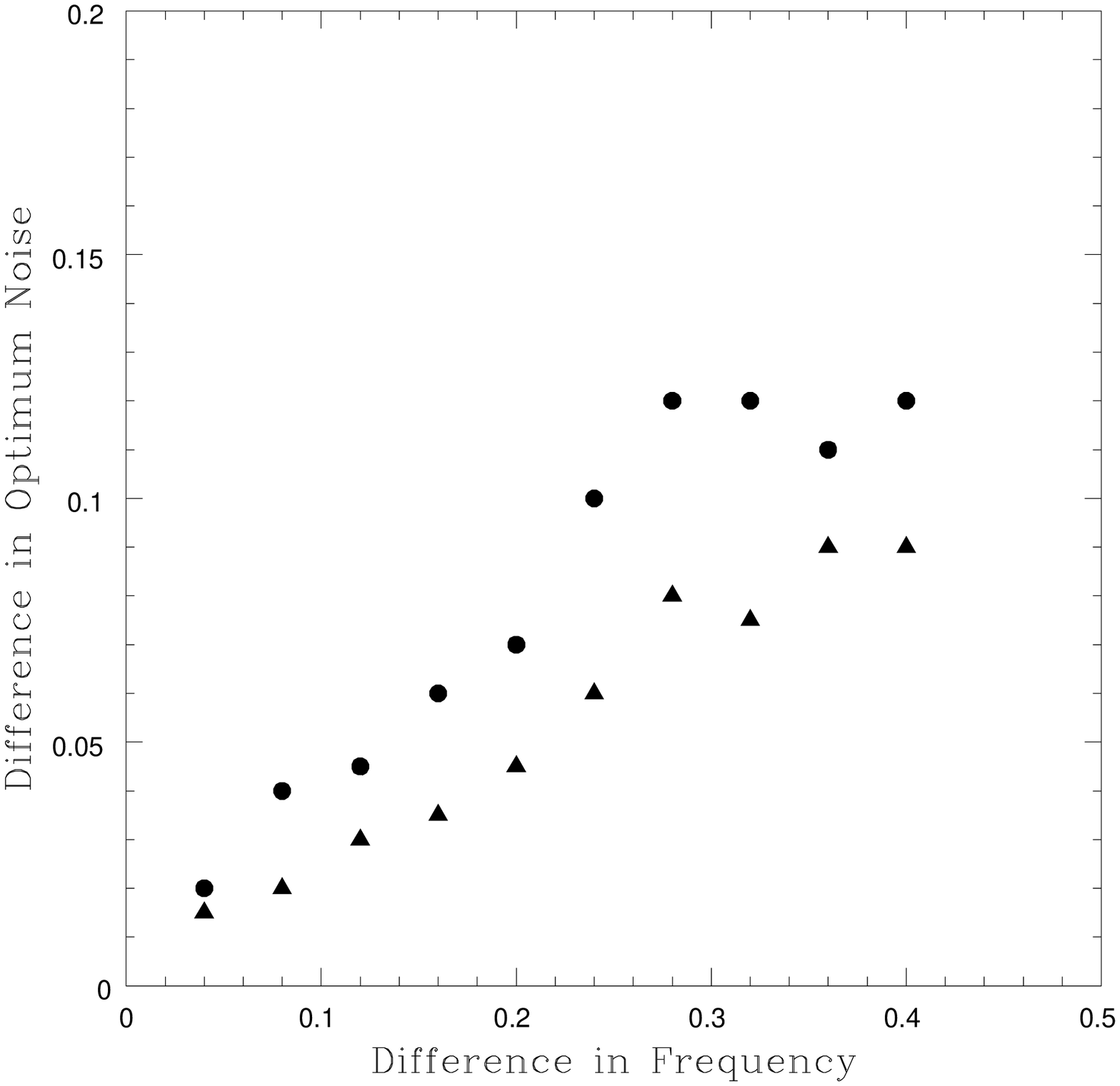}}}
\end{center}
\caption{The difference in the noise amplitudes $\Delta E$ corresponding to 
optimum SNRs plotted as a function of the difference in the input 
frequencies $\Delta \nu$, for a signal consisting of 2 frequencies. Variation 
is almost similar both for additive noise(filled circles) and 
multiplicative noise(filled triangles).}
\label{Fig.12label}
\end{figure}

This result can be understood as follows: When two signals of frequencies 
$\nu_{1}$ and $\nu_{2}$ and equal amplitude $Z$ are superposed, the resulting 
signal consists of peaks of amplitude $2Z$ repeating with a frequency 
$(\nu_{1} - \nu_{2})/2$ in accordance with the linear superposition 
principle:
\begin{equation}
Sin(2 \pi \nu_{1} t) + Sin(2 \pi \nu_{2} t) = 2Sin(2 \pi \nu_{+}t)Cos(2 \pi \nu_{-}t)
\label{eqn13}
\end{equation}
where $\nu_{+}=(\nu_{1}+\nu_{2})/2$ and $\nu_{-}=(\nu_{1}-\nu_{2})/2$ as 
explicitely shown in Fig.2. For a 
threshold system, the probability of escape depends only on the amplitude of 
the signal which is maximum corresponding to the frequency $\nu_{-}$. But 
in the case of a bistable system, the signal is enhanced only if there is a
regular shuttling between the wells at the corresponding frequency. This is 
rather difficult for the frequency $\nu_{-}$ because, its amplitude is 
modulated by a higher frequency $\nu_{+}$. This result has been checked by 
using different combinations of frequencies $(\nu_{1},\nu_{2})$ and also with 
different amplitudes. It should be mentioned here that this result reveals a 
fundamental difference between the two mechanisms of SR and is independant of 
the model considered here. We have obtained identical results with a 
fundamentally different model showing SR, namely, a model for Josephson 
junction and has been discussed elsewhere[15].

\section{Results and discussion}
In this paper we undertake a detailed numerical study of SR with a 
bichromatic input signal and gaussian white noise, in both the bistable and 
threshold set ups. We use a simple model for this purpose and both additive 
and multiplicative noise are used for driving the system in the bistable 
set up. Our analysis reveal some fundamental differences between the two 
mechanisms of SR with respect to amplification of multisignal inputs. In particular,  
we find that, while the bistable set up responds only to the fundamental frequencies 
present in the input signal, the threshold mechanism enhances a mixed mode also.

An interesting result we have obtained with a bistable system is that the SNR 
of a signal $\nu_1$ can be improved in general by adding a second signal 
$\nu_2$, of higher frequency. This is evident from fig.7 for 
multiplicative noise. It shows certain co-operative behavior between the two 
signals. To get a better understanding of the phenomenon, we plot in Fig.11 
the optimum SNR value of the first frequency $\nu_1=0.125$ for a range of 
values of $\nu_2$, for additive as well as multiplicative noise. It is clear 
that the presence of a second signal of higher frequency enhances the signal 
detectability of the first one by improving its SNR. Moreover, if the second 
signal is of the same frequency, the SNR is improved considerably indicating 
some resonance like phenomenon. Similar results, of using a high frequency 
driving for improving the detection of a low frequency signal, have been 
presented earlier [14,15] under specific dynamical set ups. But the 
simplicity of our model suggests that the result could be true in general.

Another important result that has emerged from our numerical studies is the 
possibility of using SR as a \emph{filter} for the detection or selective 
transmission of the fundamental frequencies in a composite signal using a 
bistable nonlinear medium and tuning the noise amplitude. For example, it can 
be seen from Fig.4 and 7 that the noise amplitudes for the optimum SNR for 
the two frequencies are different. The difference $\Delta E$ varies with the 
difference in the frequencies, $|\nu_1-\nu_2| (\equiv \Delta \nu)$, as shown 
in Fig.12 for both types of noise. Note that here the amplitudes of the two 
signals are equal, $Z$. The difference $\Delta E$ can be made to increase 
further by tuning the signal amplitudes suitably. Moreover, if the amplitude 
of one component goes below a minimum threshold value, the SR does not occur 
and the corresponding component is suppressed in the output. This suggests 
that SR can, in principle, be used as an effective tool for signal 
detection/transmission in noisy environments. This can be achieved basically 
in two ways, either by tuning the noise amplitude or by tuning the 
amplitude/frequency of the input signals for systems with fixed noise 
background. A similar idea has been proposed recently [12] in connection 
with the signal propagation along a one dimensional chain of coupled 
overdamped oscillators. There it was shown that noise can be used to select 
the harmonic components propagated with higher efficiency along the chain.

In the threshold mechanism of SR with a composite input 
signal, a frequency absent in the input signal is enhanced in the system 
response, a result that is not possible in the context of linear signal 
processing. This also could have potential practical applications, especially 
in the study of neuronal mechanism underlying the detection of pitch of 
complex tones[13,26]. Thus the two mechanisms of SR are different from a 
practical point of view as well and requires a more detailed analysis.

\vspace{12pt}
\ack{This work was carried out under project No.SP/S2/E-13/97(PRU) spnsored by 
Dept. of Science \& Technology, Govt.of India. The authors thank the 
hospitality and computing facilities in IUCAA, Pune.}

\end{document}